\documentclass{aa}
\usepackage{subfigure} %
\usepackage{graphicx}

\usepackage{txfonts}
\usepackage{natbib}
\usepackage{verbatim}

\usepackage{hyperref}
\hypersetup{
    colorlinks = true,
    citecolor ={blue}
}

\usepackage[pdftex]{color}
\bibpunct{(}{)}{;}{a}{}{,} 

\usepackage{booktabs}
\definecolor{ultramarine}{rgb}{0.07, 0.1, 0.6} 
\definecolor{myblue}{rgb}{0.07, 0.2, 0.6} 
\definecolor{dopal}{rgb}{.70, .25, .05}

\begin{document}

%
\title{New DA white dwarf models for asteroseismology of ZZ Ceti stars}


\author{Leandro G. Althaus\inst{1,2} and   Alejandro H. C\'orsico\inst{1,2} }
\institute{Grupo de Evoluci\'on Estelar y Pulsaciones. 
           Facultad de Ciencias Astron\'omicas y Geof\'{\i}sicas, 
           Universidad Nacional de La Plata, 
           Paseo del Bosque s/n, 1900 
           La Plata, 
           Argentina
           \and
           IALP - CONICET, La Plata, Argentina
           }
\date{Received ; accepted }

\abstract{Asteroseismology is a powerful tool to infer the evolutionary status and chemical 
stratification of white dwarf stars, and to gain insight into the physical processes that 
lead to their formation. This is particularly true for the variable hydrogen-rich atmosphere (DA) 
white dwarfs, known as DAV or ZZ Ceti stars. They constitute the most numerous class of pulsating 
white dwarfs.}
{We present a new grid of white dwarf models that take into account the advances in the last decade in the 
modeling and input physics of both the progenitor and the white dwarf stars, thus avoiding and improving 
several shortcomings present in the set of white dwarf models employed in the asteroseismological analyses 
of ZZ Ceti stars we carried out in our previous works.}
{We generate white dwarf stellar models appropriate for ZZ Ceti  stars with masses from $\sim 0.52$ to  $\sim 0.83 M_{\sun}$,  resulting from the whole evolution of initially 1.5 to  $4.0 M_{\sun}$ 
mass star models. These new models are derived from a self-consistent way with the changes 
in the internal chemical distribution that result from the mixing of all the core
chemical components induced by mean molecular-weight inversions, from $^{22}$Ne diffusion,  Coulomb sedimentation, and from residual nuclear burning. In addition, the expected nuclear-burning history and mixing events along the progenitor evolution are accounted for, in particular the occurrence of third dredge-up, which determines the properties of the core and envelope of post-AGB and white dwarf stars, as well as the white dwarf initial-final mass relation. The range of hydrogen envelopes of our new ZZ models extends from the maximun residual hydrogen content predicted by the progenitor history, $\log(M_{\rm H}/M_{\sun})\sim -4$ to $-5$,  to  $\log(M_{\rm H}/M_{\sun})= -13.5$, thus allowing, for the first time,
to make available stellar models that would enable to find seismological solutions for ZZ Ceti stars with extremely thin hydrogen envelopes, if they exist in nature. We compute adiabatic gravity($g$)-mode pulsation periods of these models. Calculations of our new evolutionary and pulsational ZZ Ceti 
models are performed with the {\tt LPCODE} stellar evolution code and the {\tt LP-PUL} stellar pulsation code.}
{Our new hydrogen-burning post-AGB models  predict chemical structures for ZZ Ceti stars substantially different from those we used in our previous works, particularly in connection with the chemical profiles of oxygen and carbon near the stellar centre. We also discuss the implications of these new models for the  pulsational spectrum of ZZ Ceti stars. Specifically, we find that the pulsation periods of  $g$ modes and the mode-trapping properties  of the new models differ significantly from those characterizing the ZZ Ceti models of our previous works, particularly for long periods.}
{The improvements in the modeling of ZZ Ceti stars we present here lead to  substantial differences in the predicted pulsational properties of ZZ Ceti stars, which are  expected to impact the asteroseismological inferences of these stars. This is extremely relevant in view of the abundant amount of photometric data from current and future space missions resulting in discoveries of numerous ZZ Ceti stars.}

\titlerunning{New white dwarf model database for asteroseismology of ZZ Ceti stars}
\authorrunning{Althaus et al.}

\maketitle

\section{Introduction}
\label{introduction}

Most stars exhibit pulsational instabilities at some stage of their evolution. 
Asteroseismology constitutes a powerful and unique tool that allows to extract
key information about  their internal structure  by comparing the independent 
periods present in their pulsation spectrum with those predicted by theoretical 
pulsation models \citep{1989nos..book.....U,2015pust.book.....C,2021RvMP...93a5001A,
2022arXiv220111629K}. In the field of white dwarf (WD) stars, this technique has 
proved to be extremely useful to infer  the evolutionary status and the internal chemical stratification of these stars \citep{2019A&ARv..27....7C}. In particular, WDs with hydrogen(H)-rich atmospheres (DA spectroscopic class) 
exhibit pulsational instabilities in a narrow strip with effective 
temperatures $10\,400$ K $\lesssim T_{\rm  eff} \lesssim  12\,400$ K  
and surface gravities $7.5  \lesssim \log  g  \lesssim 9.1$. These 
pulsating DA WDs, or ZZ Ceti stars (also called DAVs), constitute 
the most numerous class of pulsating WDs and show multi-periodic luminosity  variations  up to 0.40 mag due to  nonradial $g$(gravity)-mode  pulsations of low harmonic degree ($\ell= 1,2$) and periods between $\sim 70$ s and $\sim 1500$ s \citep[see][for reviews]{2008ARA&A..46..157W,2008PASP..120.1043F,2010A&ARv..18..471A,2019A&ARv..27....7C}.

Two main avenues are usually considered in asteroseismic modeling of 
pulsating WD stars.  One approach assumes WD structures 
characterized by parameterized chemical composition profiles. 
This constitutes a powerful forward method that allows a comprehensive  exploration of the parameter space  to infer an  optimum
asteroseismological solution \citep[see][and references therein]{1998ApJS..116..307B,2001ApJ...552..326B,2002ApJ...573..803M,2008ApJ...675.1505B,2014ApJ...794...39B,2019ApJ...871...13B,2009MNRAS.396.1709C, 2013MNRAS.429.1585F, 2016MNRAS.461.4059B,2017A&A...598A.109G,2017ApJ...834..136G,2021arXiv210615701G}. The other approach, 
which rests on chemical profiles predicted by the complete evolutionary history of progenitor stars, is a grid-based approach developed by La Plata Group\footnote{\tt http://evolgroup.fcaglp.unlp.edu.ar/} and applied 
to GW Vir stars \cite[pulsating PG 1159 stars;][]{2008A&A...478..869C}, V777 Her stars \citep[pulsating helium(He)-rich atmospheres or DBVs;][]{2012A&A...541A..42C}, ELMV stars \citep{2017A&A...607A..33C}, and ZZ Ceti stars \citep[][hereinafter R12]{2012MNRAS.420.1462R}. The asteroseismological fits we carried out in R12 for 44 bright ZZ Ceti stars and subsequent works, like \cite{2013ApJ...779...58R} and \cite{2017ApJ...851...60R}, are the first ones based on a grid of fully evolutionary ZZ Ceti models characterized by consistent chemical profiles from the center to 
the surface and covering a wide range of stellar masses, thicknesses of the H envelope 
and effective temperatures. Specifically, such studies used the grid of DA WD models developed in \cite{2010ApJ...717..897A} and \cite{2010ApJ...717..183R}. One of the main emphasis of such models is the evolutionary history of WD progenitors, since the internal
chemical stratification of WDs is the result of numerous
processes that take place during progenitor evolution, basically core He burning and  convective boundary mixing during this stage,
and the whole Asymptotic Giant Branch (AGB) and thermally pulsing AGB (TP-AGB) phases. The impact of these processes on the pulsational properties of WDs has recently been shown to be important and should be taken into account in detailed
asteroseismological analysis of pulsating WDs \citep{2017A&A...599A..21D,2018A&A...613A..46D}. 

Uninterrupted observations from current and future space missions like  
{\sl TESS} \citep[][]{2015JATIS...1a4003R}, {\sl Cheops} \citep[][]{2018A&A...620A.203M}, and {\sl PLATO} \citep[][]{2018EPSC...12..969P} 
together with {\sl Gaia} \citep{2016A&A...595A...1G} data
are expected to dramatically increase the number of ZZ Ceti stars 
\citep[see][]{2022MNRAS.511.1574R}, providing  unprecedented high-quality observations \citep{2020FrASS...7...47C,2022arXiv220303769C}. 
The interpretation of these data demands the development of a new generation
of fully evolutionary ZZ Ceti models. In this paper, we present a new 
grid  of WD models appropriate for conducting future asteroseismological studies of ZZ Ceti stars, with the main emphasis on the impact of their chemical profiles on the Brunt-V\"ais\"al\"a frequency and pulsational adiabatic periods.
This new grid of WD models constitutes a major improvement over the 
ZZ Ceti models developed by \cite{2010ApJ...717..897A} and 
\cite{2010ApJ...717..183R} 
used in the asteroseismological fits in R12. Specifically, we concentrate on generating 
WD stellar models with masses in the range of
$\sim 0.52$ to  $\sim 0.83 M_{\sun}$, resulting from the whole evolution
of initially 1.5 to  4.0 $M_{\sun}$ mass star models,
which embraces the range of stellar masses expected for most of the observed ZZ Ceti stars. The new grid presented here  takes into account
the advances in the last decade  in the modeling and input physics of both the progenitor and WD stars that will be necessary to exploit the potential of
the incoming observations of ZZ Ceti stars.

The paper is organized as follows. In  Sect.~\ref{models} we  describe  the major improvements of our
new grid of models over that  used in R12. 
In Sect.~\ref{pulsation_results}  we
describe the consequences for the pulsational properties
of ZZ Ceti stars.  Finally,  in Sect.~\ref{conclusions}  we summarize  the main findings of the paper.

\section{The new WD evolutionary models}
\label{models}

\begin{table*}[t]
\centering
  \caption{Main characteristics of our progenitor and WD sequences.    
    $M_i$: Initial mass of the model (at ZAMS). $M_{\rm WD}$: Final WD mass. HeCF: Full He-core flash at the 
    beginning of the core He-burning phase. $\tau_{\rm pre-WDMS}$: Lifetime from
    the ZAMS to the onset of WD cooling branch.
    $M_c^{1TP}$: Mass of the H-free core at the
    first thermal pulse (defined as those regions with $X_{\rm
      H}<10^{-4}$). $ M_{\rm H}$: Mass of the H content at the maximum effective temperature
      at the beginning of the WD cooling branch. $M_{\rm H}^{\rm ZZ}$: Mass of the H content at the ZZ Ceti instability strip. $N_{\rm C}/N_{\rm O}:$ C/O ratio in number fraction at the end of
    the TP-AGB phase.}
\begin{tabular}{cccccccc} 
  \hline
  \hline             
  $M_{\rm i}$ & $M_{\rm WD}$  & HeCF & $\tau_{\rm pre-WD}$  & $M_{\rm c}^{\rm 1TP}$ &  log $M_{\rm H}$ & log $M_{\rm H}^{\rm ZZ}$ & $N_{\rm C}/N_{\rm O}$\\
  $[M_\odot]$ &   $[M_\odot]$ & & [Myr] &  $[M_\odot]$  &    $[M_\odot]$   & $[M_\odot]$ &    \\
  \hline
  1.00     &   0.5281  & yes  & 11813      &   0.5119     &   $-3.60$   &   $-3.93$ &   0.40  \\
  1.25     &   0.5614  & yes  & 5256.6     &   0.5268     &   $-3.77$   &   $-4.04$ &   0.37  \\
  1.50     &   0.5759  & yes  & 2820.6     &   0.5267     &   $-3.93$   &   $-4.28$ &   2.20  \\
  2.00     &   0.5803  & no   & 1418.9     &   0.4873     &   $-4.11$   &   $-4.23$ &   1.12  \\
  3.00     &   0.6573  & no   & 442.92     &   0.6103     &   $-4.47$   &   $-4.59$ &   1.66  \\
  4.00     &   0.8328  & no   & 195.90     &   0.8086     &   $-5.17$   &   $-5.32$ &   0.81  \\
  \hline  
  \\
  \end{tabular}                 
    \label{tab:TPAGB}
\end{table*}

\begin{figure}
        \centering
        \includegraphics[width=1.0\columnwidth]{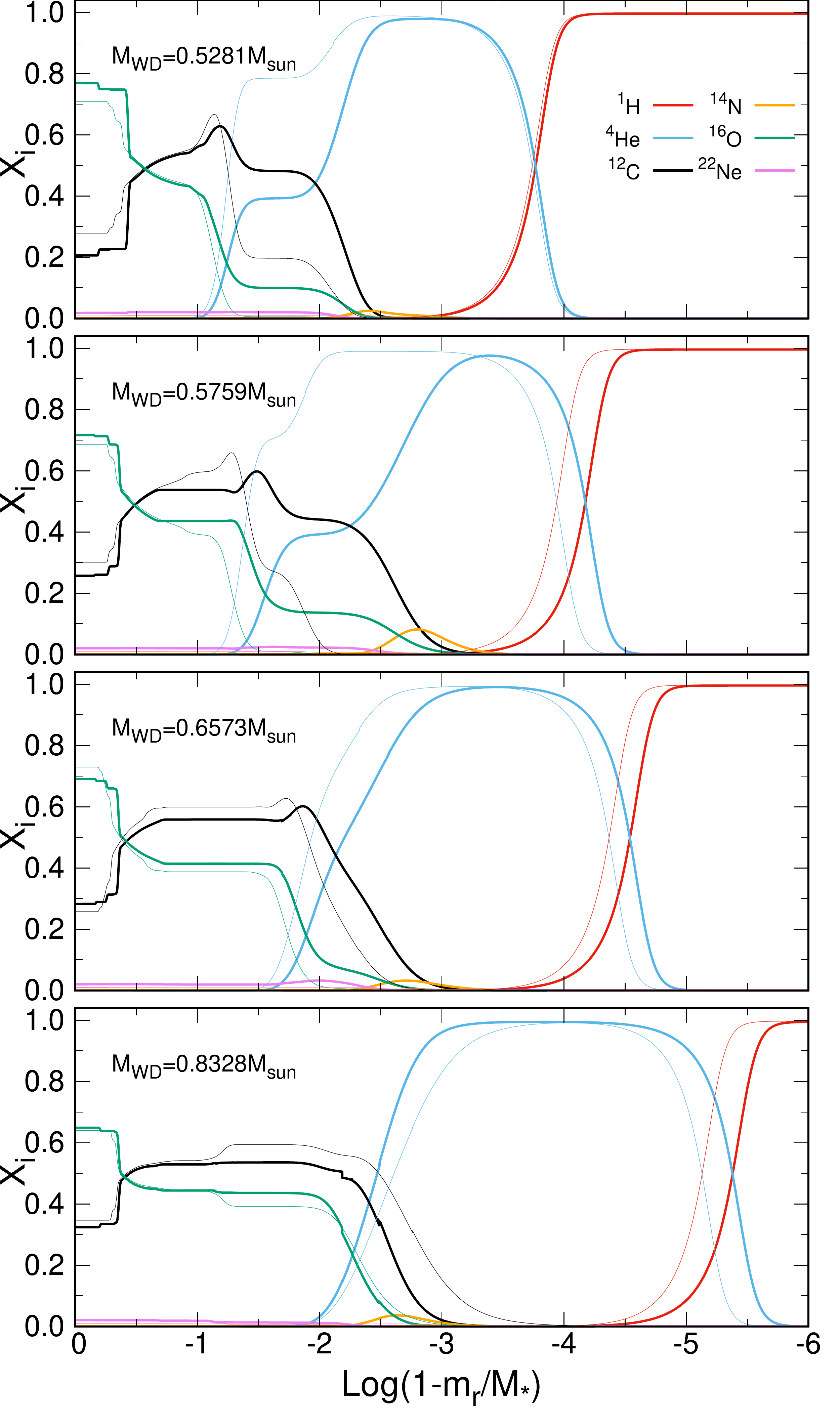}
        \caption{Abundance by mass of $^1$H, $^4$He, $^{12}$C, $^{14}$N,
          $^{16}$O, and $^{22}$Ne in terms of the outer mass coordinate  for our 
           0.5281, 0.5759, 0.6573, and 0.8328$M_{\sun}$ WD  models at
           the domain of the pulsating ZZ Ceti stars. Our new chemical profiles
           are compared with those we used in R12 of similar stellar masses (thick
          and thin lines, respectively).} 
        \label{perfiles_12000.eps}
\end{figure}

The new models presented here avoid several shortcomings that are present in the set of evolutionary ZZ Ceti models developed in \cite{2010ApJ...717..897A}, set that has constituted the base for the asteroseismological fits we carried out in R12 and subsequent works. More importantly, the new models take into account
new advances in the modeling and input physics of both the progenitor and
WD stars that will be of utmost importance for future asteroseismological inferences of WDs.  

The new set of WD models uses as an input  the complete evolution of progenitor stars evolved from the main sequence through the TP-AGB phase, by employing a recently updated version of our stellar evolution code  {\tt LPCODE} developed by La Plata group. {\tt LPCODE} provides WD models characterized by  consistent
chemical profiles for both the core and envelope as a result of the full
computation of the evolution of progenitor stars. 
This code has  been widely used and tested in numerous  stellar evolution contexts of low-mass and  WD stars \citep[see][for details]
{2003A&A...404..593A,2005A&A...435..631A, 2013A&A...555A..96S, 2015A&A...576A...9A,
2016A&A...588A..25M,2020A&A...635A.164S,
2020A&A...635A.165C}. In particular, and relevant
for this work, {\tt  LPCODE} computes WD evolution
in a self-consistent way with the changes in the internal chemical distribution that result from the mixing of  all the core chemical components induced by  mean molecular  weight  inversions left
by prior evolution, element  diffusion, and convective mixing. Also,  residual nuclear burning 
down to advanced evolutionary stages of WD evolution is
allowed. The role played by residual H burning
is not negligible \cite[see][for recent observational evidence of quiescent thermonuclear activity occurring in cooling WDs]{2021NatAs...5.1170C}  and by the time evolution reaches the ZZ Ceti domain, it has yielded  a substantial reduction of the H content with which the WD enters its cooling branch. 

We highlight below the improvements of our
new set of models over those presented in \cite{2010ApJ...717..897A}
that are expected to impact the period spectrum of ZZ Ceti stars:

\begin{enumerate}

\item {\it Thickness of H envelope}: the range of H envelope has 
been extended to $\log(M_{\rm H}/M_{\sun})=-13.5$, in contrast with 
the lowest H content we adopted in R12 of $\log(M_{\rm H}/M_{\sun})=-9.5$. 
This will allow to find seismological solutions for ZZ Ceti stars with extremely thin H envelopes (if they exist). This was not possible in R12.
To this end, we have artificially reduced 
the H content of our post-
AGB sequences, so as to mimick a more efficient H mixing to deeper layers and burning expected during a very late thermal pulse (VLTP) episode, a possible
single evolution scenario that predicts the formation of H-poor WDs, \citep[see][for a recent discussion]{2019A&ARv..27....7C}. As a result of this procedure,
only tiny vestiges of H remain, extending from the surface down to deeper layers. As the WD cools down, such traces of H float to the surface as a result of gravitational settling, thus  forming an increasing pure (and thin) H envelope. To properly follow this process, {\tt LPCODE} considers a new fully implicit treatment of time-dependent
element  diffusion that  includes thermal, chemical  diffusion,
gravitational settling \citep{2020A&A...633A..20A}, and Coulomb diffusion (see below).  We follow the diffusion of the isotopes  $^1$H,$^3$He, $^4$He, $^{12}$C,  $^{13}$C,
$^{14}$N,$^{15}$N, $^{16}$O, $^{17}$O,$^{18}$O, $^{19}$F,  $^{20}$Ne, and $^{22}$Ne.

\item {\it Treatment of core He burning}: the new set of models presented here are based on a more realistic implementation of convective boundary mixing toward the end of He-core burning \citep{2016A&A...588A..25M} than what we performed in R12. This yields a larger central oxygen (O) abundance for the same initial mass $M_i$ and a very abrupt variation of the abundances of O and carbon (C) in the core.

\item {\it $^{22}$Ne diffusion}: because of the two neutron excess of $^{22}$Ne, this isotope rapidly
sediments in the interior  of  WDs \citep{1992A&A...257..534B}, thus releasing an additional amount of energy that delays the cooling times  \citep{2002ApJ...580.1077D,2010ApJ...719..612A,2010Natur.465..194G}. More recently, \cite{2016ApJ...823..158C} have explored   the impact  of $^{22}$Ne  sedimentation on  the adiabatic  pulsational properties  of ZZ~Ceti models and found that this process induces appreciable changes in the pulsation periods of ZZ Ceti stars (in the range $10-50$ s depending on the stellar mass) that have to be taken into account in  attempts to perform precise  asteroseismology of these stars. A similar conclusion has recently been arrived at
in \cite{2021ApJ...910...24C} in the case of DBV variable stars, who find
the presence of a systematic offset in the periods of pulsating WD when $^{22}$Ne diffusion is absent. In our new set of WD models, $^{22}$Ne diffusion has been considered from initial realistic $^{22}$Ne abundances in the WD interior, resulting from  the full evolution of solar metallicity progenitor stars.
This constitutes a major improvement over our seismological models employed in R12, particularly in the case of more massive WDs. 

\item {\it Coulomb sedimentation}: another piece of WD physics that was neglected in R12
was the separation of ions due to Coulomb interactions at high densities. This process affects the diffusion flux in WDs \citep{2013PhRvL.111p1101B,2010ApJ...723..719C}, driving  the  gravitational settling of ions with the same mass to charge number $A/Z$, as in the case
of plasmas  rich in $^{4}$He and  $^{12}$C. Recently,
\cite{2020A&A...644A..55A} have shown that, in the case of massive WDs, 
Coulomb separation produces  marked changes in the $g$-mode pulsation periods 
(up to $\sim 15$ \% ) that cannot be neglected  in   detailed asteroseismological 
analyses  of ZZ Ceti stars. Coulomb sedimentation has been incorporated in a time-dependent
diffusion treatment, as given in \cite{2020A&A...644A..55A}. 

\item {\it Progenitor evolution}: the evolutionary history of progenitor star determines
the internal chemical profile of the WD, and  hence the pulsational properties of ZZ Ceti stars.  The starting configurations for our new set of WD models are the H-burning  post-AGB  star  models with metallicity $Z=0.02$   derived  by \cite{2016A&A...588A..25M}, who considered  new observational constraints and recent advances in the micro-  and macro-physics involved in the core He burning, AGB  and TP-AGB  modeling. Among these improvements
we mention the inclusion of updated low-temperature opacities for the
C- and O-rich AGB stars,  a
consistent treatment of the stellar winds for the C- and O-rich
regimes, and an updated and improved description of convective boundary mixing processes during the thermal pulses and
previous evolutionary stages\footnote{Convective boundary mixing during
the thermal pulses was disregarded in the computation of progenitor stars of the
ZZ Ceti models used in R12.}. These improvements lead to  stellar models characterized by
the occurrence of appreciable third dredge-up (3DUP) and C
enrichment of the envelope, that reproduce
many relevant AGB and post-AGB observables, in particular the C/O ratios of AGB and post-AGB stars in the Galactic disk, the WD initial-final mass
relation (IFMR) at near solar metallicities, the C-O
intershell abundances of AGB stars as observed in PG1159 stars, and
the mass range of C-rich stars in the stellar clusters of the
Magellanic Clouds. In addition, the properties of the resulting AGB stellar models such as core growth, C enrichment, and IFMR are  well in agreement with
the predictions of modern AGB grids, so the predicted structures  can be considered good representatives of state-of-the-art AGB
stellar evolution modeling for the post-AGB phase \citep[see][for details]{2016A&A...588A..25M}. 3DUP episodes, which are strongly inhibited in 
progenitor stars of the ZZ Ceti models used in R12, alter the IFMR, thus impacting the core chemical composition and the envelope
structure of a WD of given mass \citep[see][]{2010ApJ...717..897A}. 
The H-burning  post-AGB sequences of \cite{2016A&A...588A..25M} also reproduce the observed range of He-, C-, and O-intershell abundances of AGB stars, as determined from
the observations of PG1159 stars \citep[see][]{2006PASP..118..183W}. This is of utmost importance for the envelope chemical stratification of ZZ Ceti stars and their pulsational properties \citep{2010ApJ...717..897A}. In addition, 3DUP episodes
lead to the formation of WDs with less massive H envelopes and
pockets of abundant $^{13}$C and  $^{14}$N. Finally, the agreement with
observational data of the C/O ratio of PNe shows that the
new models depart from the TP-AGB at the right time in terms of
C enrichment, thus giving confidence in the accuracy of the 
thermochemical structure of our WD models \citep[see][]{2016A&A...588A..25M}. 

\end{enumerate}

Table  \ref{tab:TPAGB} summarizes some characteristics of progenitor stars of our new ZZ Ceti models that are relevant for the WD formation. The initial mass ($M_{\rm i}$) of progenitor stars
at the ZAMS ranges from 1.5 to  4.0 $M_{\sun}$,   which yield final WD masses ($M_{\rm WD }$) in the range 0.5281 to  0.8328 $M_{\sun}$. Some features deserve some comments. We note that the 1.5 to  2.0 $M_{\sun}$ progenitors lead to
quite similar WD masses, inflicting a pronounced plateau in the IFMR at that
range of masses. This is connected with the transition from degenerate to nondegenerate He-core ignition at that range of initial masses and the resulting different sizes of the H-free core during He-core burning. This explains the minimum in the H-free core mass at the first thermal pulse (see column 5 in Table  \ref{tab:TPAGB} at $M_{\rm i} \approx$ 2 $M_{\sun}$).

Most of our sequences experience efficient 3DUP episodes and
C enrichment of the envelope, with the exception of the lower mass models.
As a result, most of our models become C-rich during the TP-AGB evolution, i.e., $N_{\rm C}/N_{\rm O}>1$ (see column 8 in  Table \ref{tab:TPAGB}), in agreement with the expected C
   enrichment as observed in many AGB and post-AGB stars \citep[see][]{2016A&A...588A..25M}. The high C/O ratio that characterizes our $M_i=1.5 M_\odot$ model results from the fact that this model underwent a 
final thermal pulse shortly after the end of the
TP-AGB. As a result,  the C dredged up to the surface is diluted
into a significantly  smaller mass of H, leading to a  higher
surface C abundance. The occurrence of 3DUP is key in determining
the IFMR. Intense 3DUP delays the growth of the H-free core during the
TP-AGB evolution, yielding smaller final WD masses  than in the
absence of 3DUP for the same initial mass $M_i$. As mentioned, this 
is relevant for the internal chemical composition of a WD of given mass. In addition, the higher C enrichment
of the envelope resulting from efficient 3DUP leads to smaller
masses of the H envelope at the departure from the AGB.  An updated treatment of the 
previous evolution, in  particular, opacities and nuclear reaction rates, also lead to models that depart from the AGB with smaller H-envelope masses and brighter luminosities \cite[see][for details]{2016A&A...588A..25M}. The masses of the H content listed in Table  \ref{tab:TPAGB} (column 6) are upper limits for the  maximum H content expected from our single progenitors at
the onset of the cooling branch. By the time the  domain of the ZZ Ceti stars
is reached, the H content has been markedly reduced by residual H burning (see coulumn 7 in Table  \ref{tab:TPAGB}).

\begin{figure}
        \centering
        \includegraphics[width=1.0\columnwidth]{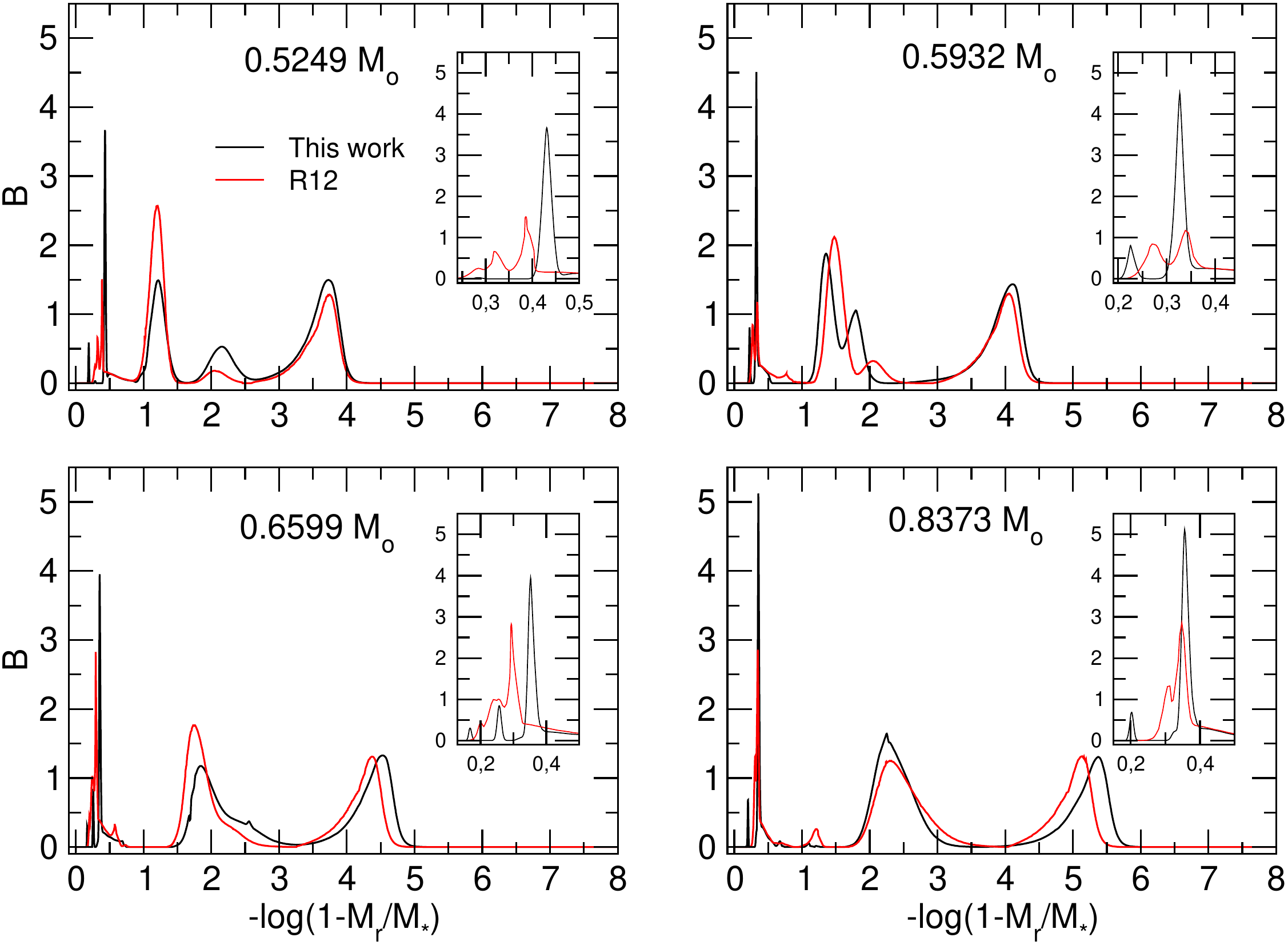}
        \caption{The Ledoux term $B$ in terms of the outer mass fraction as 
        predicted by our DA WD template models with different masses and 
        a thick H envelope  
        and by those used in R12 (black and red lines, respectively). The various features exhibited by $B$ can be directly related to the different chemical transition regions of the models (see Fig. \ref{perfiles_12000.eps}). In each panel, the inset is a zoom of the C/O-core region.} 
        \label{b_ledoux}
\end{figure}

 \begin{figure}
        \centering
        \includegraphics[width=1.0\columnwidth]{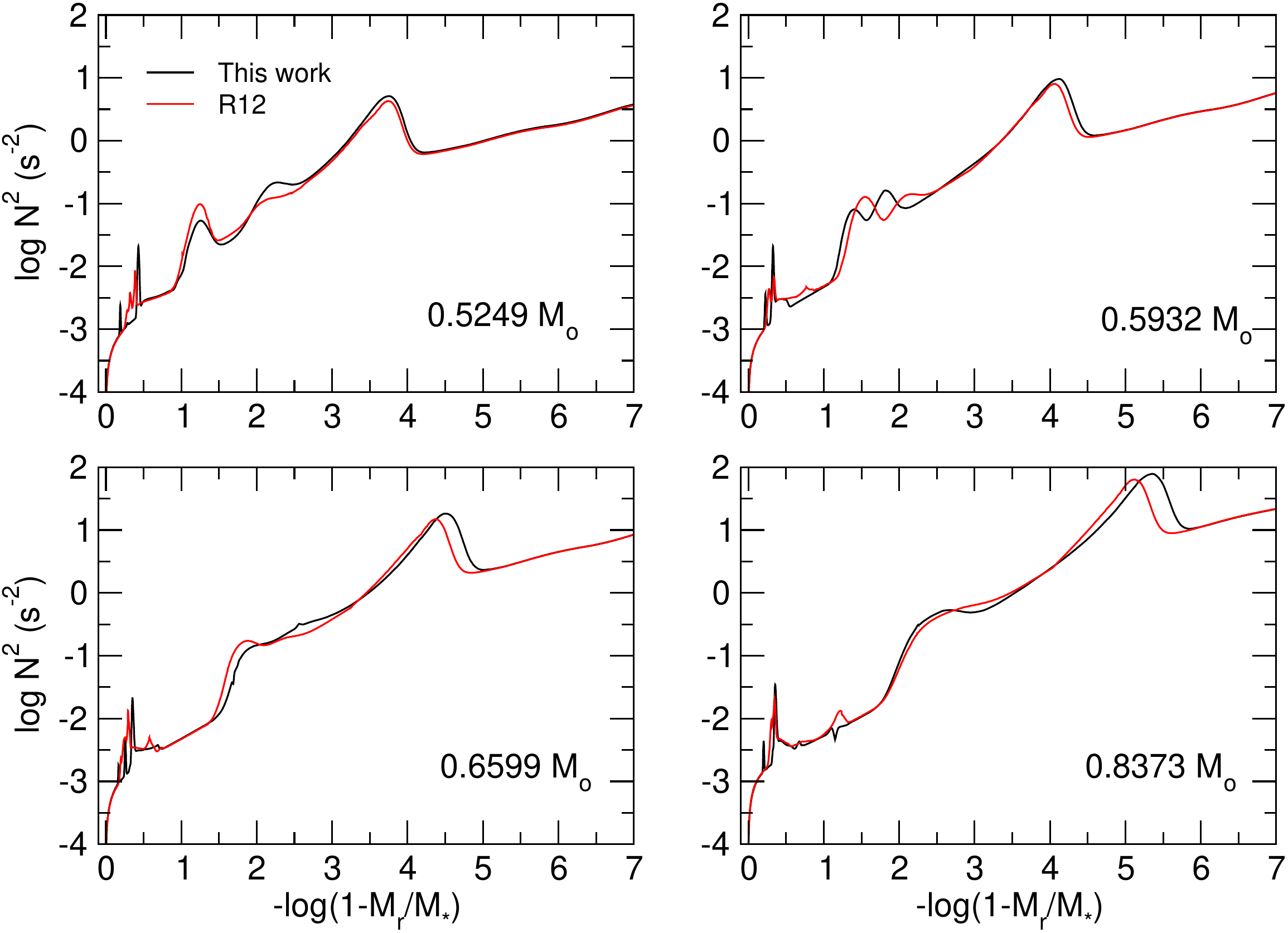}
        \caption{Same as Fig. \ref{b_ledoux}, but for the Brunt-V\"ais\"al\"a frequency. } 
        \label{bvf}
\end{figure}

\begin{figure*}
        \centering
\includegraphics[width=1.7\columnwidth]{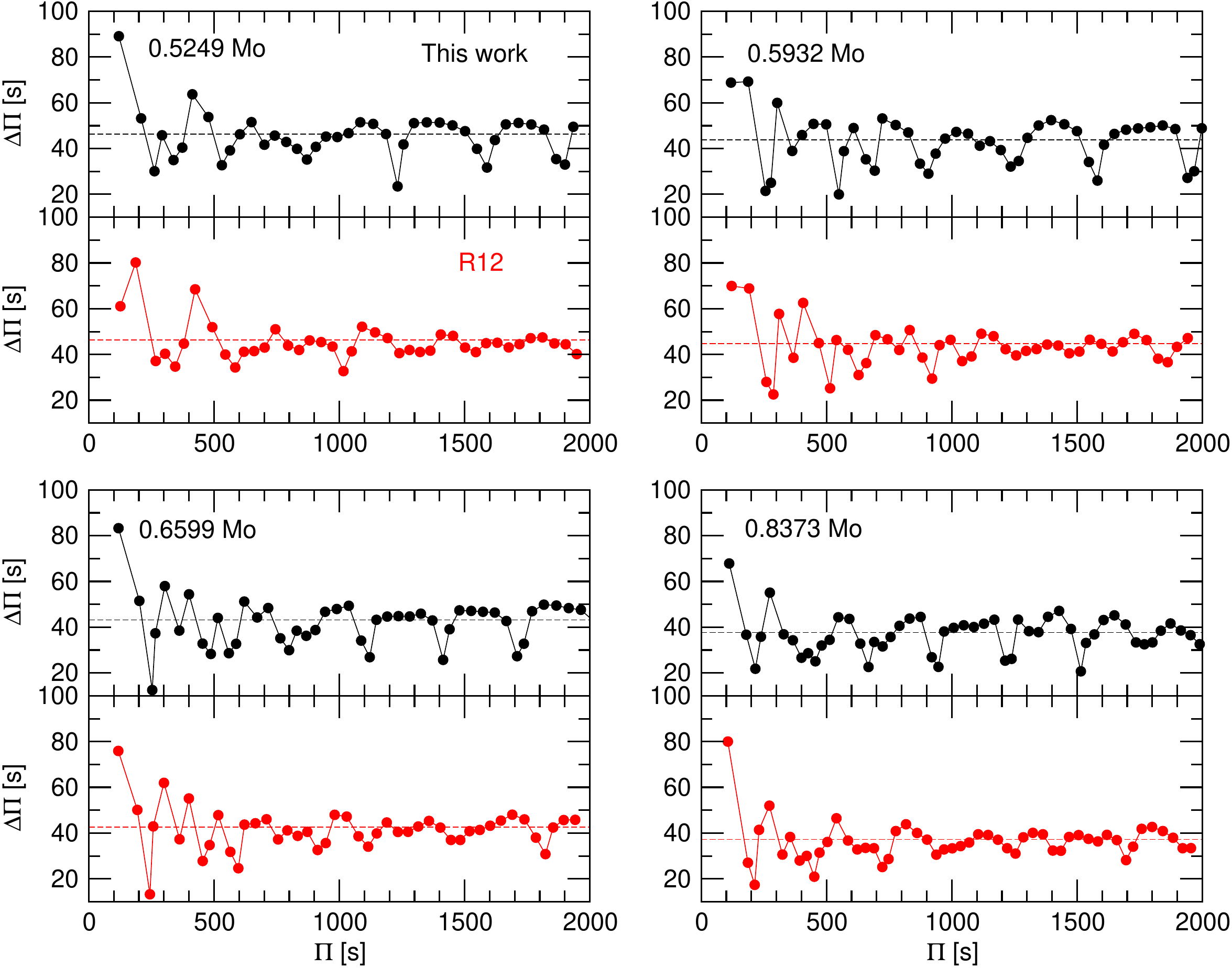}
        \caption{The dipole ($\ell= 1$) forward period spacing versus periods for 
        our template DA WD models with thick H envelopes, say $\log(M_{\rm H}/M_{\star})$ 
        between $-4$ and $-5$,  and for the models 
        of R12 with similar H thicknesses (black dots  and red dots, respectively). The asymptotic period
        spacing for both sets of models is plotted with dashed lines.} 
\label{period_spacing}
\end{figure*}

\subsection{Implications for the chemical profiles}
\label{chemical} 

The starting WD configurations provided by our new H-burning
post-AGB models  were evolved  from the planetary nebulae stages 
down to the domain of the ZZ Ceti stars in  a consistent  way with abundance  changes resulting 
from   element  diffusion, convective mixing,  and residual nuclear  burning. At high effective temperatures, we performed  the  core mixing  implied  by  the  inversion of  the  mean
molecular  weight  left by  prior  evolution.
The resulting chemical stratification for some selected isotopes by the time
evolution   reached   the   ZZ  Ceti domain, at $T_{\rm eff} \sim 12\,000$ K, is displayed in Fig. \ref{perfiles_12000.eps}
for the  0.5281, 0.5759, 0.6573, and 0.8328$M_{\sun}$ WD  models, and compared
with the chemical stratification of ZZ Ceti models  used in R12 of similar stellar masses (thick
and thin lines, respectively)

\begin{figure}
        \centering
\includegraphics[width=1.0\columnwidth]{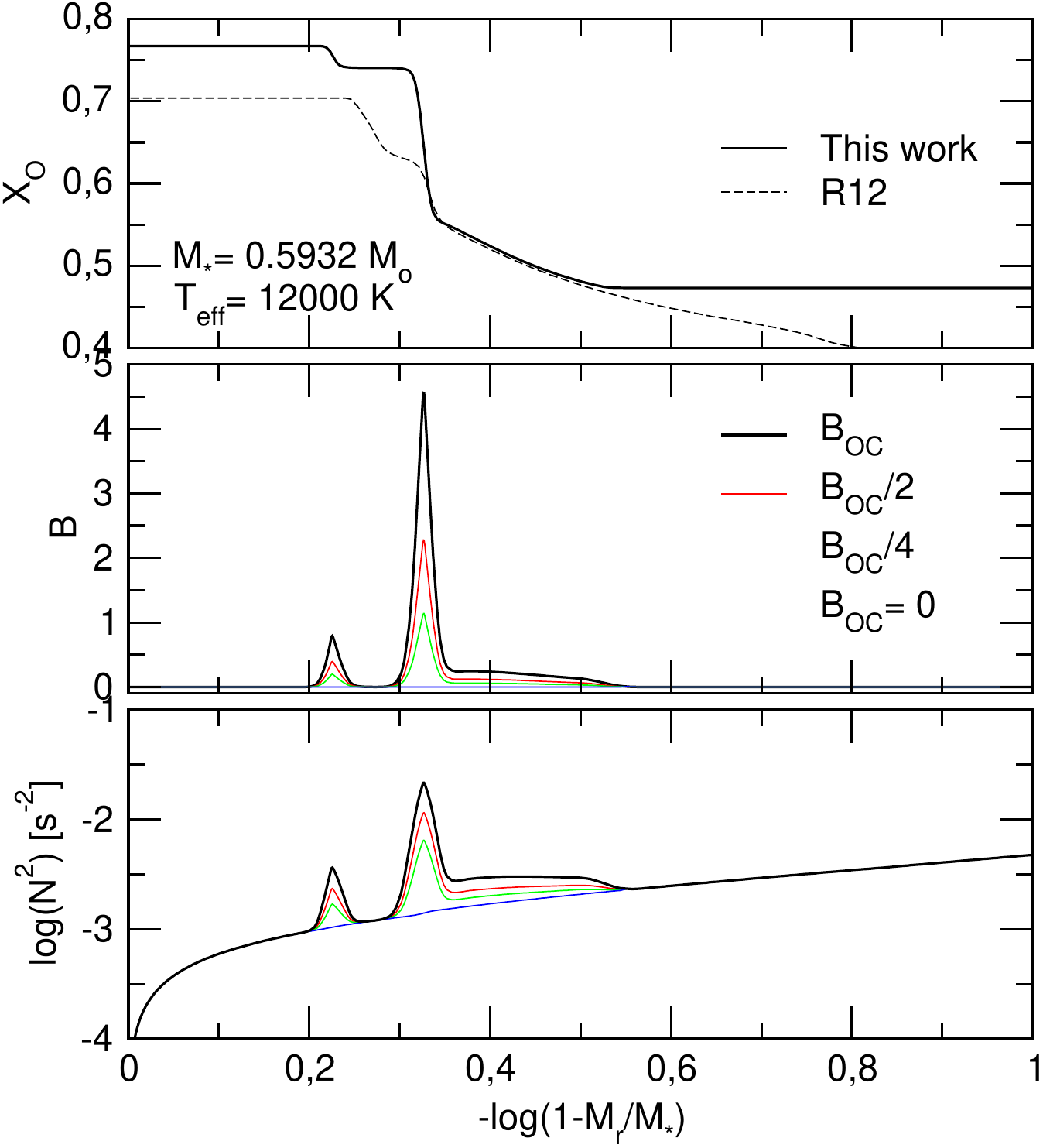}
        \caption{{\sl Upper panel:} the chemical profile of O at the CO-core regions of a 
    DA WD model with $M_{\star}= 0.932 M_{\odot}$, $T_{\rm eff}\sim 12\,000$ K, and a thick H envelope, for the case of the present calculations (solid black line) and the R12 computations (dashed black line). {\sl Middle panel:} the Ledoux term $B$ at the region of the O/C interface,  $B_{\rm OC}$, of the same model depicted in the upper panel, along with the 
    artificial situations of $B_{\rm OC}/2$, $B_{\rm OC}/4$, and $B_{\rm OC}= 0$. 
    {\sl Lower panel:} the logarithm of the squared Brunt-V\"ais\"al\"a frequency for the same 
    cases considered in the middle panel.} 
\label{x-b-n2-bvariable}
\end{figure}

\begin{figure*}
        \centering
\includegraphics[width=1.7\columnwidth]{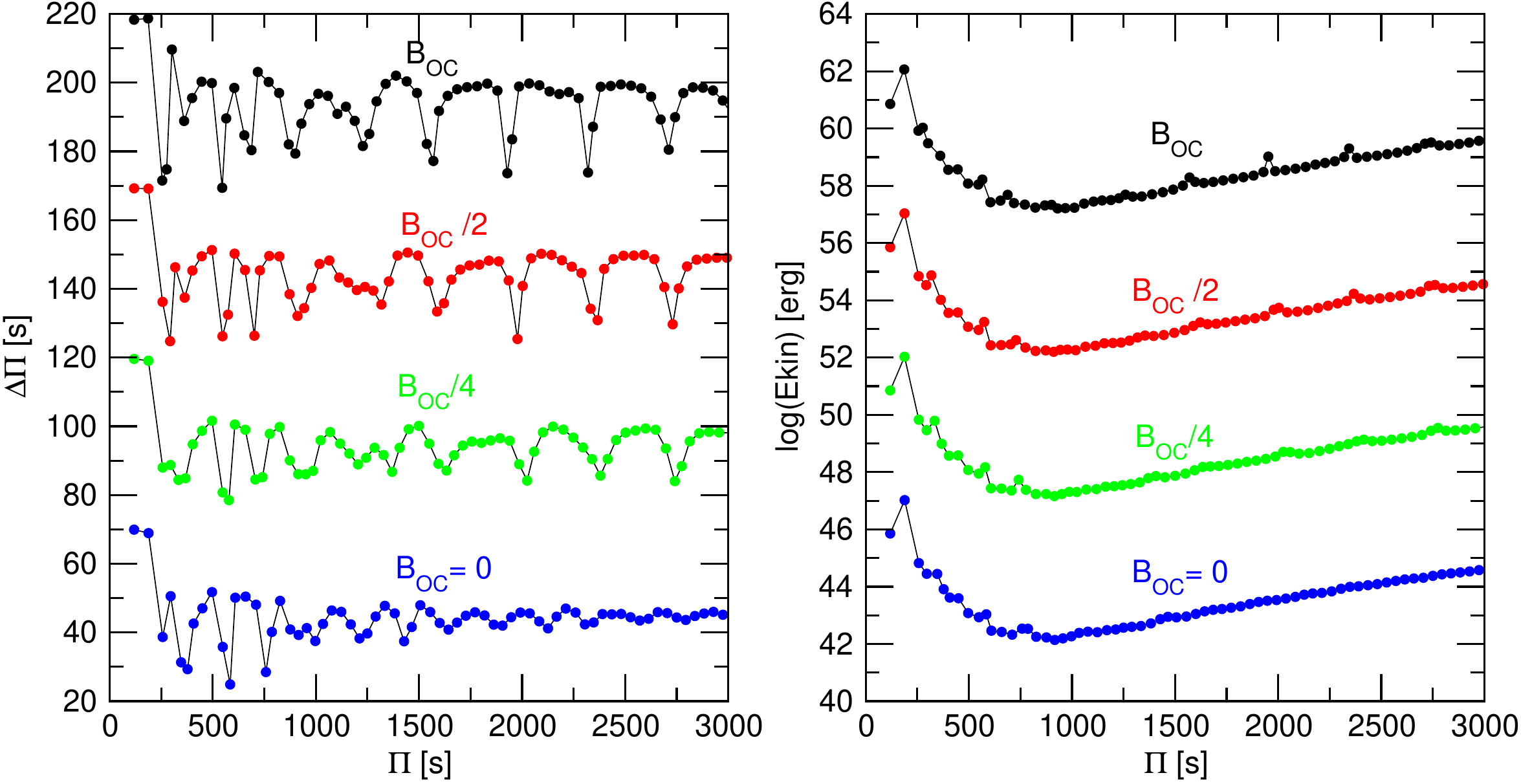}
        \caption{{\sl Left panel}: the $\ell= 1$ forward period spacing $\Delta \Pi$ in terms of the 
        pulsation periods, for decreasing  magnitude of Ledoux term $B$ at the CO chemical 
        interface ($B_{\rm OC}$), from the 
        top to the bottom panel. The curves have been artificially displaced up for clarity, where $y$-axis values make sense only for the case $B_{\rm OC}= 0$. {\sl Right panel:} similar to the left panel, but for the oscillation kinetic energy. The models 
        have $T_{\rm eff}\sim 12\,000$ K, $M_{\star}= 0.5932 M_{\odot}$, and thick H envelopes.}
\label{b_variable}
\end{figure*}

\begin{figure}
        \centering
\includegraphics[width=1.0\columnwidth]{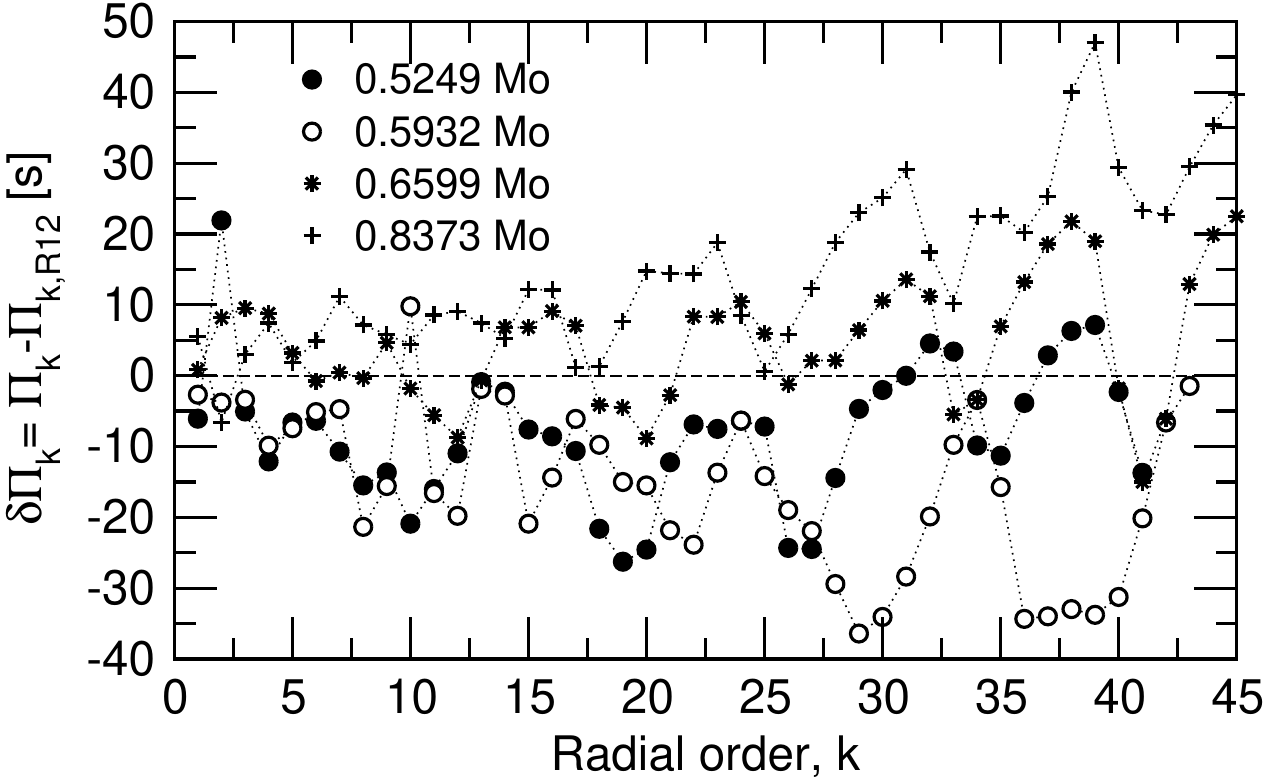}
        \caption{Differences between the $\ell= 1$ pulsation periods of our template DA WD models with thick H envelope and the periods corresponding to the R12 models, in terms of the radial order, for different stellar masses.} 
\label{deltaperiod}
\end{figure}

Clearly, the new H-burning post-AGB models  predict chemical structures for the ZZ Ceti stars substantially different from those characterizing the R12 models. 
The internal composition of  WDs is a crucial  aspect for the determination  of the pulsational  properties of  these stars. The impact of these new chemical structures on 
the theoretical  pulsational spectrum of ZZ Ceti stars should hence be assessed. Three main
improvements in the treatment of progenitor evolution of our new ZZ Ceti models, which have a direct impact on the WD chemical profile, can be identified. First, as mentioned, the more careful implementation of convective boundary mixing toward the end of He-core burning performed in  \cite{2016A&A...588A..25M}, which yields a larger central O abundance for the same initial mass $M_i$. This feature implies a very abrupt variation of the abundances of O and C in the core, which, as we will show later (Sect. \ref{pulsation_results}), has a 
dramatic impact on the pulsational properties of our new models. Second, the occurrence of appreciable 3DUP during the TP-AGB phase in our progenitor stars, which leads to smaller WD masses, particularly for intermediate mass stars.  In particular, our new progenitor treatment predicts an initial mass of 
$M_i= 3 M_{\sun}$ for the 0.6573  \,$M_{\sun}$ WD  model, while
the same WD mass corresponds to a $M_i= 2.5M_{\sun}$
progenitor star in the set of ZZ Ceti models  used in R12. This explains the smaller central O abundance we find
for this WD model, as compared with that of R12\footnote{At this range
of  $M_i$, the central O abundance decreases with increasing $M_i$.}. Third,  the evolutionary history of WD progenitors includes the possibility that they may experience the last thermal pulse shortly after departing from the TP-AGB. This is relevant concerning the location in mass of the He-rich buffer and the underlying intershell region rich in He and C build up by the He-flash convection zone at the last AGB thermal pulse.
This is particularly clear in the  $0.5759 M_{\sun}$ WD model resulting from the $M_i= 1.5 M_{\sun}$ progenitor that experienced
a late thermal pulse. We note in this case that both the location and 
chemical profile of the intershell region result markedly affected
by the occurrence of a late thermal pulse.   The
presence of C in the intershell region stems from the mixing due to the short-lived He-flash convection zone that  pushed the C-rich  zone upward during
the peak of the last He thermal pulse in the AGB. Because of the inclusion of 
convective boundary mixing during the TP-AGB in our new sequences, particularly
at the pulse-driven convection zone, a larger
O abundance is predicted in the intershell region.
Also, the  large abundance of $^{14}$N  in the He-rich buffer,
in  particular for  the $0.5759 M_{\sun}$ ZZ Ceti model, is due to the
occurrence of  appreciable 3DUP in its progenitor star and the subsequent
H burning in a C-enriched medium.
We also note that the more efficient 3DUP experienced by our progenitor stars leads to maximum masses of the  H content that are markedly smaller than those characterizing the set of ZZ Ceti models of R12 and described in \cite{2010ApJ...717..897A} and \cite{2010ApJ...717..183R}. This is not the case for the 0.5281$M_{\sun}$ ZZ Ceti model, the progenitor of which avoids the C enrichment of the envelope (see Table  \ref{tab:TPAGB}),
resulting in a final H content similar to that of \cite{2010ApJ...717..183R}.  
Finally, for both sets of ZZ Ceti models, the H and He contents left by progenitor evolution decrease with WD mass.



During  WD cooling,  element  diffusion processes  modify the  chemical abundance  distribution in  the outer  layers  left by prior mixing
and burning episodes during the TP-AGB evolution. This is more noticeable
in the case of more massive WDs, for which the inter-shell rich
in He, C, and O is completed eroded by chemical diffusion by the time the
domain of the ZZ Ceti stars is reached. For the least massive WD
model, we note that such intershell region is still present at this stage. 
The presence of such double-layered structure affects the pulsational
periods of ZZ Ceti stars \citep[see][]{2010ApJ...717..897A}. In the more massive models, chemical diffusion is also responsible for the presence of abundant C in the He-buffer zone. Another improvement of our new ZZ Ceti
models which, as mentioned, impacts on the predicted pulsation periods
of ZZ Ceti stars, is $^{22}$Ne diffusion. Here, the abrupt change
in the initial $^{22}$Ne profile at the edge of the CO core is strongly 
smoothed out by diffusion, leading to the formation of 
a bump in its abundance.  Finally, the chemical profiles shown in Fig. \ref{perfiles_12000.eps} bear the signature of Coulomb sedimentation.
This process is relevant for the more massive WDs
and for mixtures of  ions with equal $A/Z$.  In this case,
Coulomb    separation     drives    gravitational     settling, and ions  with larger $Z$  move to
deeper layers. We note that the H-He interface is not affected
by  Coulomb diffusion because in  this  case the  contribution due  to
gravity is dominant, and Coulomb diffusion represents a minor 
contribution to the  diffusion flux. 
Both $^{22}$Ne diffusion and Coulomb sedimentation were not considered in the
set of ZZ Ceti models employed in R12.
  
\begin{figure}
        \centering
        \includegraphics[width=1.0\columnwidth]{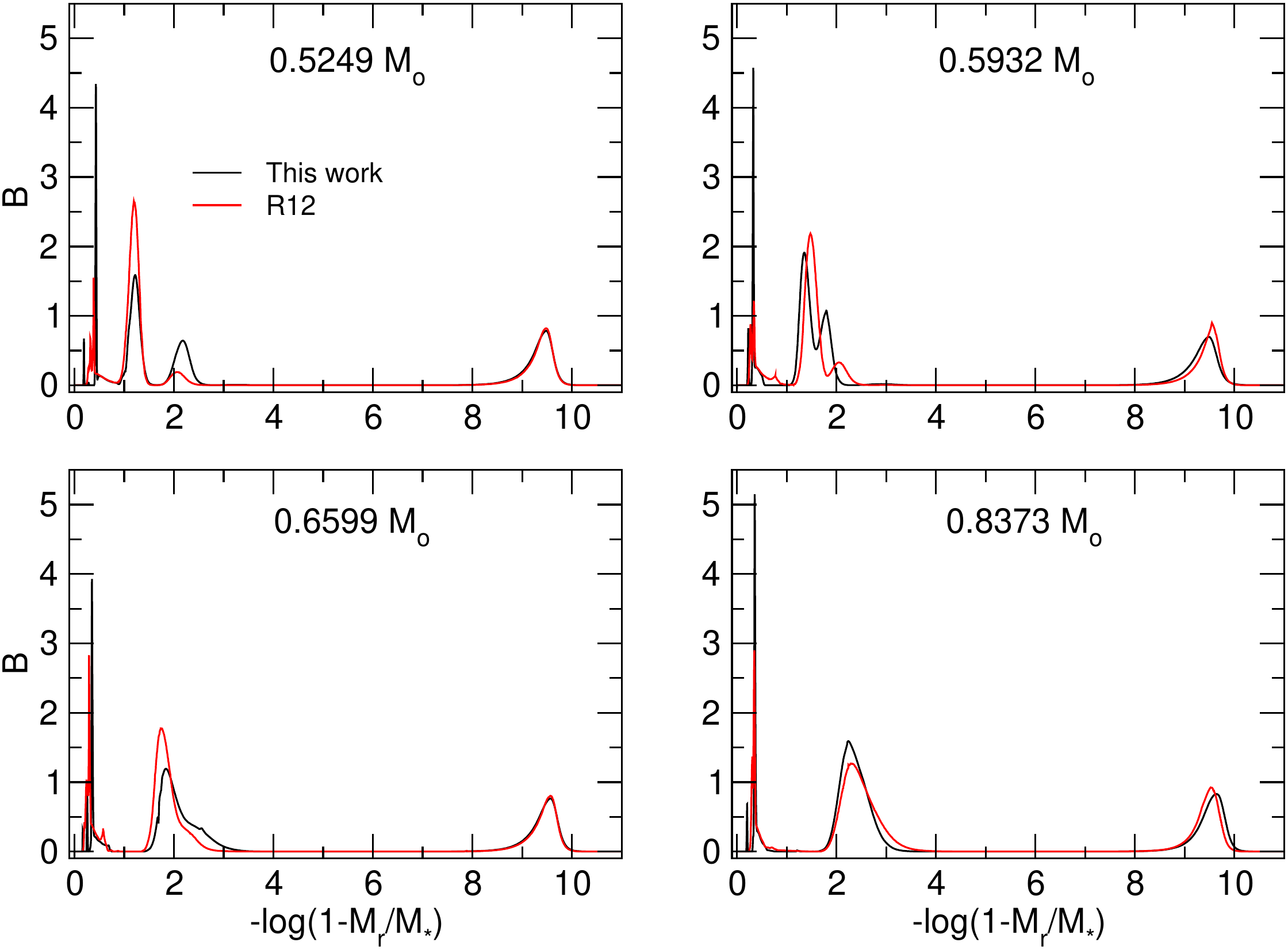}
        \caption{Same as Fig. \ref{b_ledoux}, but for DA WD template models with a thin H envelope.} 
        \label{b_ledoux-thin}
\end{figure}

\begin{figure}
        \centering
        \includegraphics[width=1.0\columnwidth]{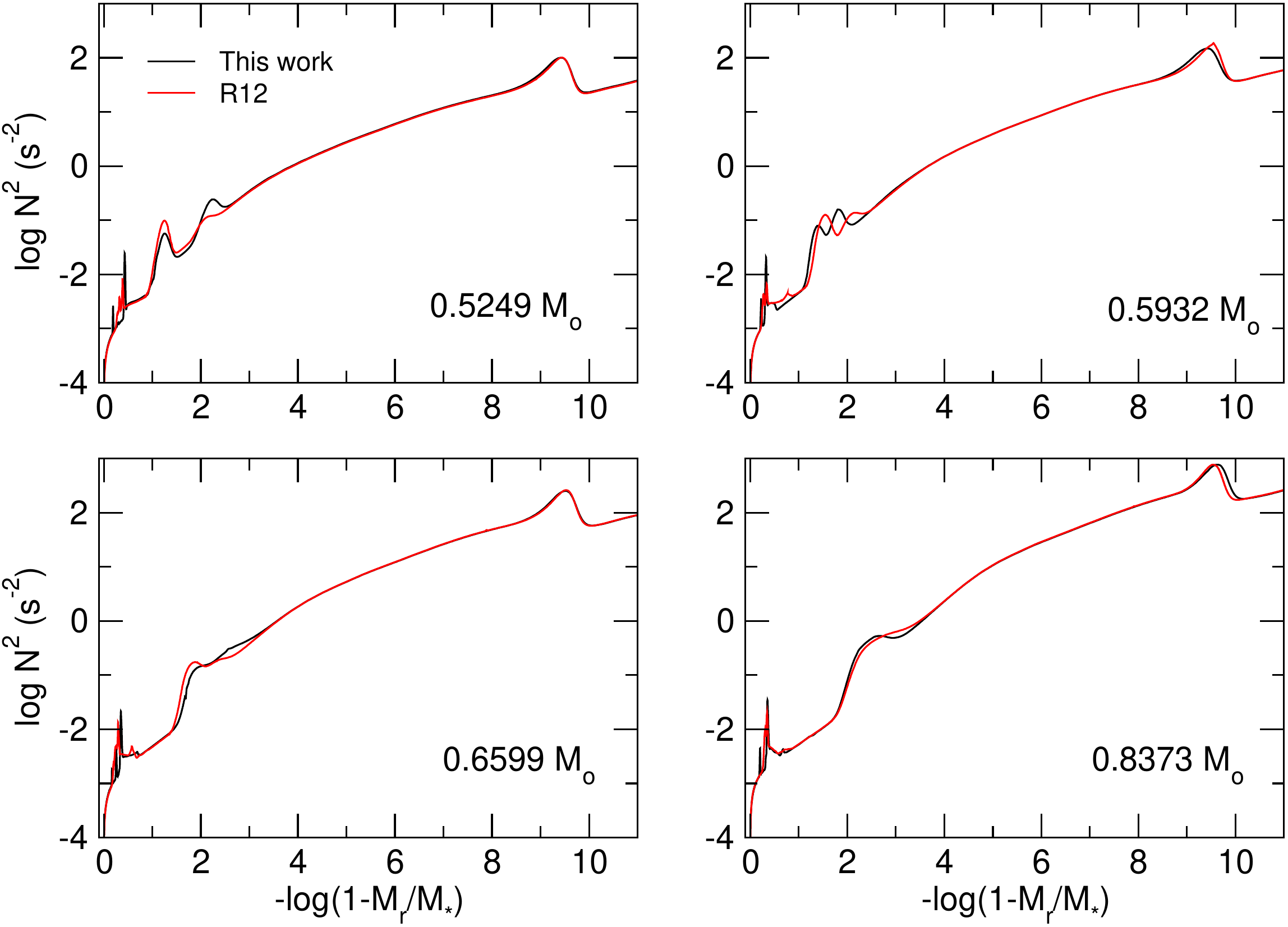}
        \caption{Same as Fig. \ref{bvf}, but for DA WD template 
        models with a thin H envelope.} 
        \label{bvf-thin}
\end{figure}

\begin{figure*}
        \centering
\includegraphics[width=1.7\columnwidth]{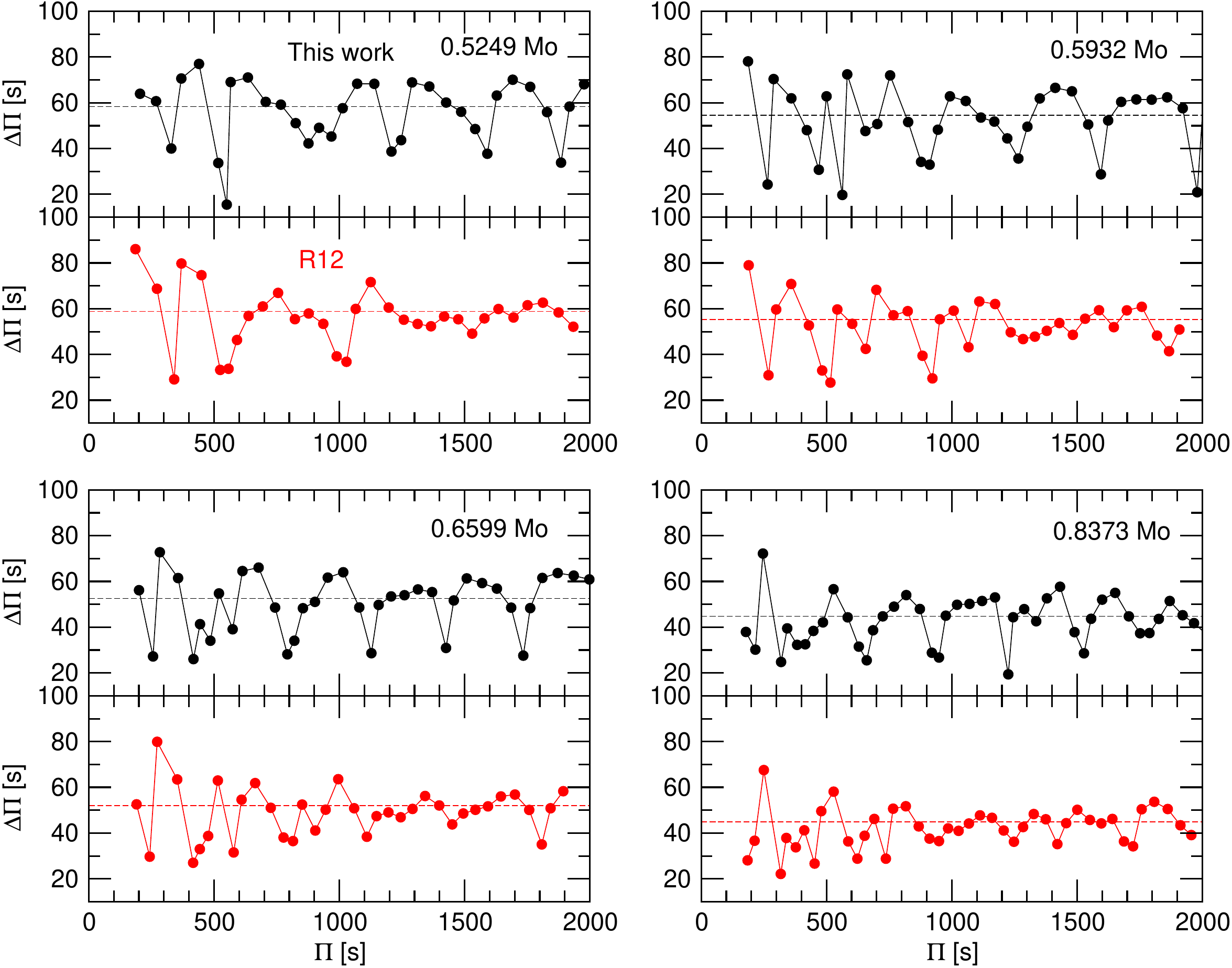}
        \caption{Same as Fig. \ref{period_spacing}, but for DA WD 
        template models with a thin H envelope, $\log(M_{\rm H}/M_{\star})\sim -9.3$.} 
\label{period_spacing-thin}
\end{figure*}

\begin{figure}
        \centering
\includegraphics[width=1.0\columnwidth]{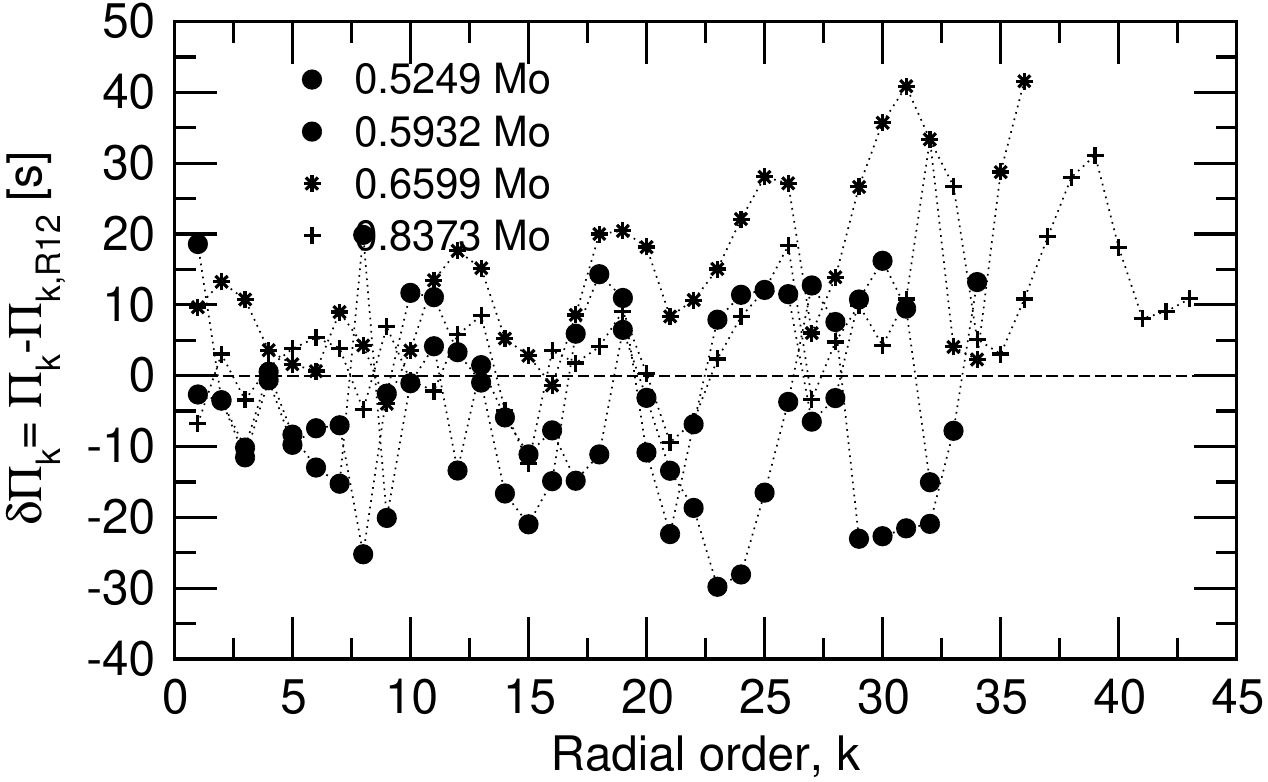}
        \caption{Same as Fig. \ref{deltaperiod}, but for DA WD 
        template models with a thin H envelope.} 
\label{deltaperiod-thin}
\end{figure}

\section{Impact on the pulsational properties of ZZ Ceti stars}
\label{pulsation_results}

In this section, we compare the pulsational properties of our new ZZ Ceti 
models with those of R12. This comparison is of utmost relevance, given that 
the R12 DA WDs models have also been generated with the LPCODE, although 
with an earlier version of it, and they are  
currently being used in the grid-based asteroseismological approach with 
fully evolutionary stellar models. Because the models of 
R12 and those that we present in this paper differ mainly in the 
internal chemical structure, we can anticipate  
that there will be appreciable differences in the oscillation 
periods and the period spacings of the $g$ modes. We emphasize
that the differences in the chemical profiles of our models as 
compared with the R12 ones are the result of the 
integrated effects of the incorporation of Ne diffusion, 
the inclusion of Coulomb sedimentation, 
the different treatment of core He burning, and the distinct 
progenitor evolution modeling. That said, here we will not examine 
the individual effect of each of these ingredients of our modeling, since 
this has been done to a great extent in other works 
\citep{2016ApJ...823..158C,2021ApJ...910...24C,2020A&A...644A..55A}. 
We will examine models with the same stellar mass and effective temperature, and 
very similar thickness of the H envelope, in order to assess the changes 
in the periods and the period spacings introduced
 by  the employment of our new evolutionary models. 

\begin{figure}
        \centering
        \includegraphics[width=1.0\columnwidth]{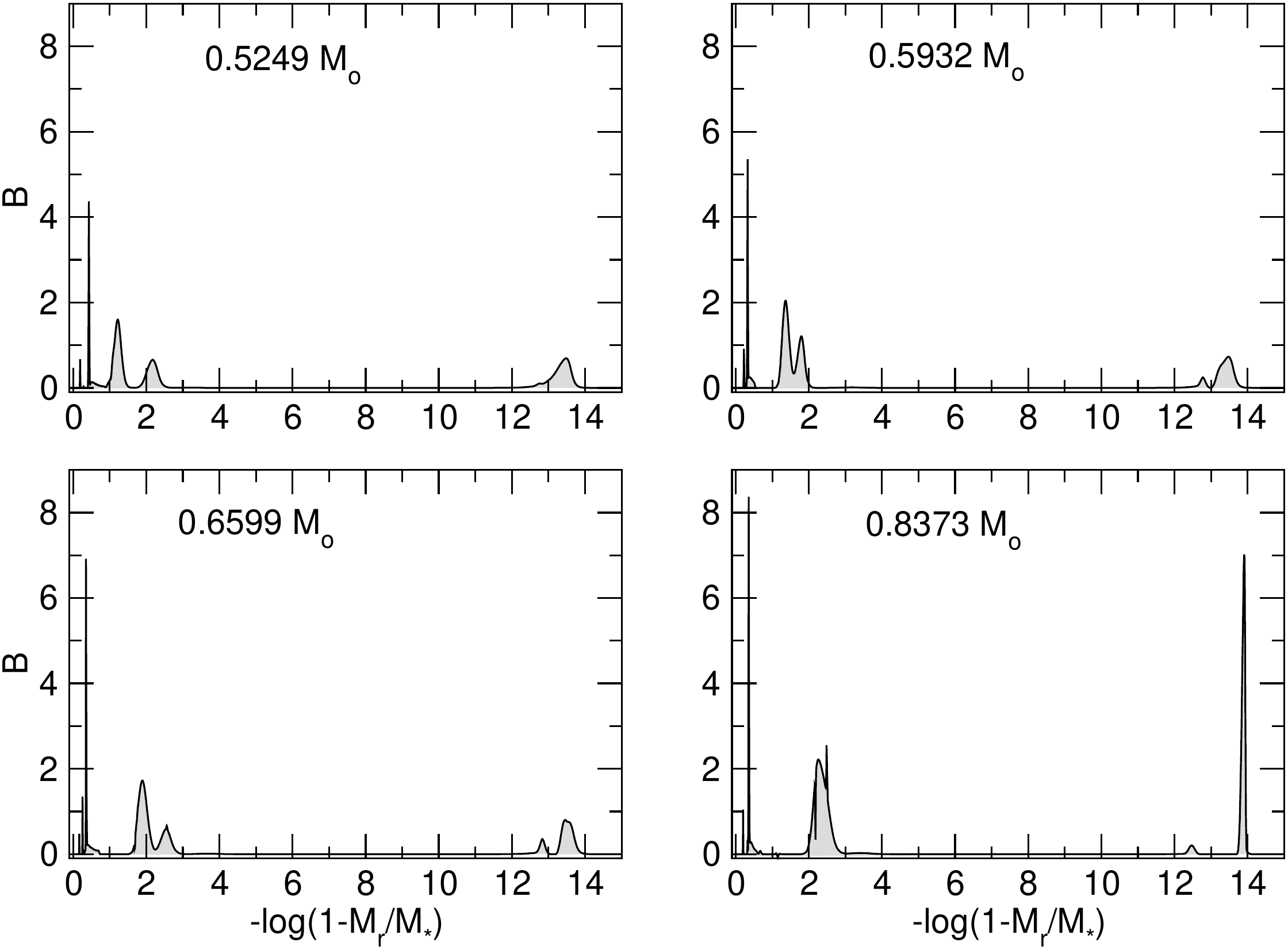}
        \caption{The Ledoux term $B$ in terms of the outer mass fraction as predicted by our DA WD template models with different masses and a ultra-thin 
        H envelope, that is $\log(M_{\rm H}/M_{\star}) \sim -13.4$.} 
        \label{b_ledoux-ultra-thin}
\end{figure}

The Ledoux term B in terms of the outer mass fraction as predicted by our DA WD template models with different masses and a thick
H envelope and by those used in R12 (black and red lines, respectively).
The various features exhibited by B can be directly related to the differ-
ent chemical transition regions of the models (see Fig. 1). In each panel,
the inset is a zoom of the C/O-core region.

 \begin{figure}
        \centering
        \includegraphics[width=1.0\columnwidth]{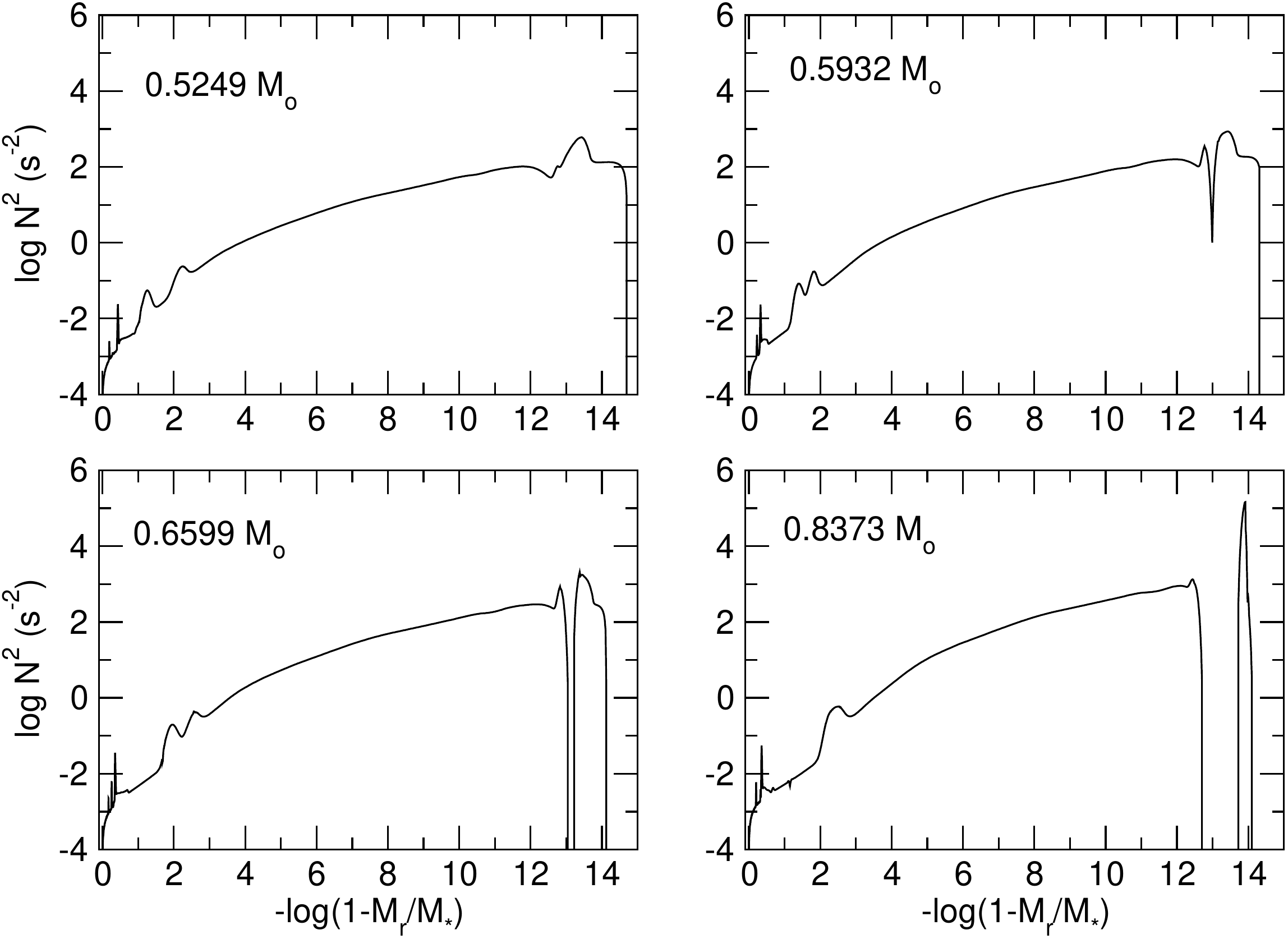}
        \caption{Same as Fig. \ref{bvf}, but for our DA WD template 
        models with a ultra-thin H envelope.} 
        \label{bvf-ultra-thin}
\end{figure}

The pulsation modes of our new ZZ Ceti models as well as those 
of the R12 models have been computed with the same adiabatic version 
of the {\tt  LP-PUL} pulsation code described in \citet{2006A&A...454..863C}. 
The squared    
Brunt-V\"ais\"al\"a  frequency ($N$, the critical frequency of nonradial 
$g$-mode pulsations) is computed as in \cite{1990ApJS...72..335T}, 
according to the following expression:

\begin{equation}
N^2= \frac{g^2 \rho}{P}\frac{\chi_{\rm T}}{\chi_{\rho}}
\left[\nabla_{\rm ad}- \nabla + B\right],
\label{bv}
\end{equation}

\noindent where $g$, $\rho$, $P$, $\nabla_{\rm ad}$ and $\nabla$ are 
the acceleration of gravity, density, pressure, adiabatic temperature 
gradient and actual temperature gradient, respectively. The 
compressibilities are defined as:

\begin{equation}
\chi_{\rho}= \left(\frac{d\ln P}{d\ln \rho}\right)_{{\rm T}, \{\rm X_i\}}\ \ \
\chi_{\rm T}= \left(\frac{d\ln P}{d\ln T}\right)_{\rho, \{\rm X_i\}}.
\end{equation}

Finally, the Ledoux term $B$ is computed as \citep{1990ApJS...72..335T}:

\begin{equation}
B= -\frac{1}{\chi_{\rm T}} \sum_1^{M-1} \chi_{\rm X_i} \frac{d\ln X_i}{d\ln P}, 
\label{BLedoux}
\end{equation}

\noindent where

\begin{equation}
\chi_{\rm X_i}= \left(\frac{d\ln P}{d\ln X_i}\right)_{\rho, {\rm T}, 
\{\rm X_{j \neq i}\}}.
\end{equation}

The  computation of  the Ledoux term includes the effects of an 
arbitrary number of chemical species that vary in abundance in the 
transition regions. In the following sections, we will first focus on examining 
DA WD models with canonical H envelopes, that is, the thickest possible H 
envelopes according to our detailed pre-WD evolutionary calculations. Then, we 
will focus on models with thin H envelopes, and finally we will concentrate 
on the pulsation properties of models with ultra-thin H envelopes.

\begin{figure*}
        \centering
\includegraphics[width=1.7\columnwidth]{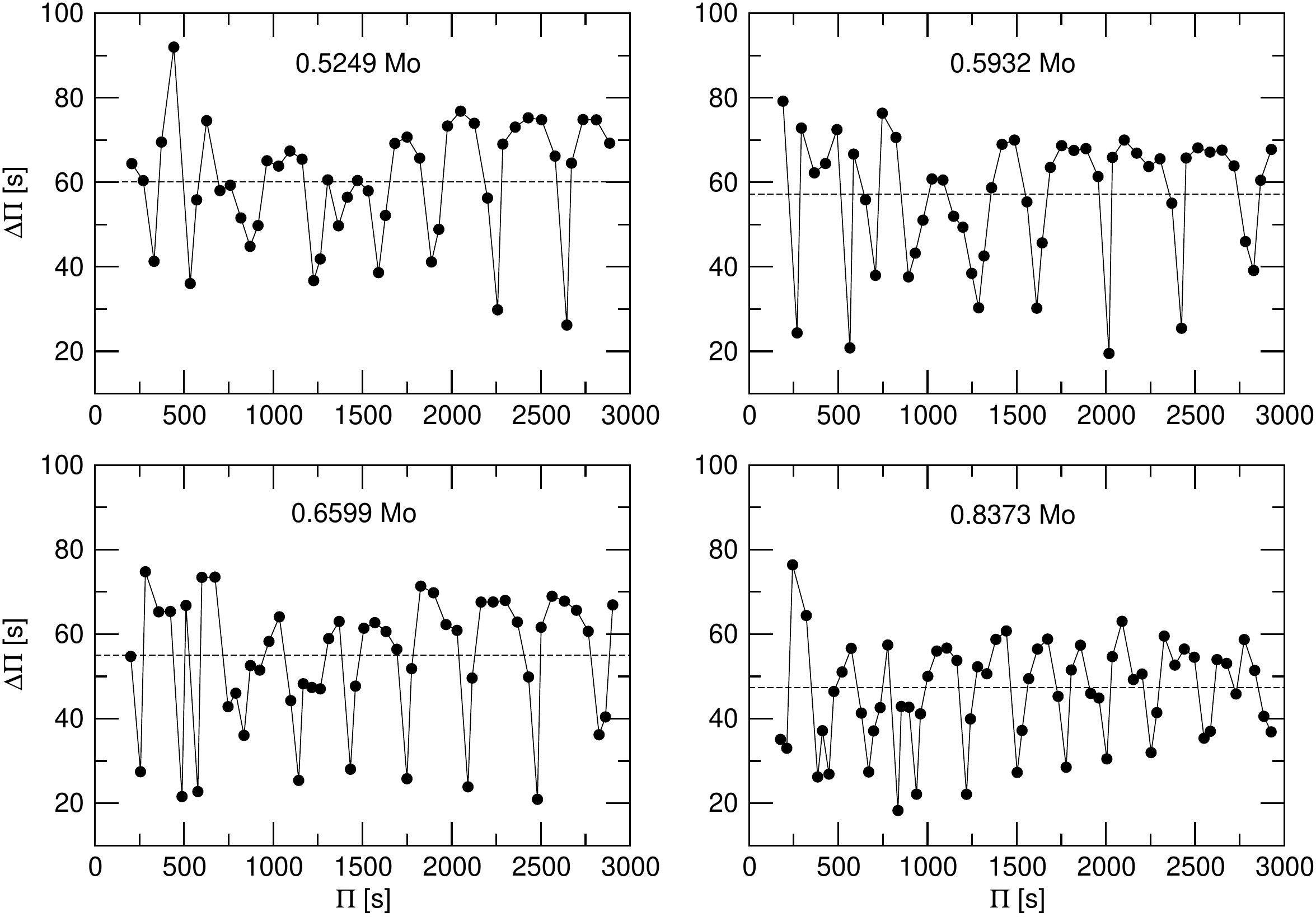}
        \caption{Same as Fig. \ref{period_spacing}, but for our DA WD 
        template models with a ultra-thin H envelope ($\log(M_{\rm H}/M_{\star}) \sim -13.4$).} 
\label{period_spacing-ultra-thin}
\end{figure*}

\subsection{DA WDs with thick H envelopes}
\label{thick}

We have considered the following values of the stellar mass: 
$0.5249, 0.5932, 0.6599$ and $0.8373 M_{\odot}$, in agreement with some sequences of
the R12 computations (their Table 1), and picked up four template models of our set
of structures with effective temperature $\sim 12\,000$ K and canonical (thick) H envelopes, $\log(M_{\rm H}/M_{\star}) \sim -4$ to $-5$. These models have been derived from interpolation from our set of models listed in Table  
\ref{tab:TPAGB}. We emphasize that, in order to isolate as much as possible 
the effects of the differences in the chemical profiles on the pulsation periods, we have adopted stellar mass values exactly equal to those considered by R12.
In Fig. \ref{b_ledoux} we show the Ledoux term $B$ 
versus the logarithm of the outer mass fraction for the template models, where we
use red curves for the R12 models. 
The insets of each panel amplify the region of the CO core. We find that the 
peaks associated with the O/C/He and He/H transition regions of our set of models are similar to those of the R12 models,  although in some cases 
($0.5249 M_{\odot}$ and $0.6599 M_{\odot}$), the peak due to 
the triple transition (O/C/He) is more important in the case of R12 models. 
The strongest difference in $B$, however, come from the contribution of 
the O/C chemical interface, which is much more  noticeable in the models 
of the present work for all the examined stellar masses. 
In particular, for the case of $M_{\star}= 0.5932 M_{\odot}$, the peak 
of the Ledoux term of our template model is more than four times higher than for the analogous model of R12. This is due to the much more abrupt step of the O and C chemical
abundances in the core of our models (see Fig. \ref{perfiles_12000.eps}), which makes the contribution to $B$ much more noticeable in the models presented in this paper than in 
the models of R12. The run of the squared Brunt-V\"ais\"al\"a 
frequency for the same models is shown in Fig. \ref{bvf}. Although the 
Brunt-V\"ais\"al\"a frequencies are almost similar in both sets of models, the peaks corresponding to the CO-core chemical transition  are  more pronounced in the new models.

The differences in $B$ and $\log(N^2)$ described above have dramatic consequences for the mode-trapping characteristics of our new models. This is revealed by plotting the 
forward period spacing ($\Delta \Pi= \Pi_{k+1}-\Pi_k$) versus the periods, as shown in Fig. \ref{period_spacing}. It is clear from the figure that, for periods of up to $\sim 900-1000$ s, the period spacing of both sets of models looks qualitatively similar, showing small differences for a few modes. However, for longer periods, the period spacing of our new models exhibits strong minima, even for very long periods ($\gtrsim 2000$ s, not plotted), while the R12 models exhibit mode-trapping features with decreasing amplitudes and with  period spacings gradually closer to the constant asymptotic limit (horizontal dotted lines), as we consider longer and longer periods. 

To find the reason for the occurrence of these minima in our models, we have 
recalculated the full spectrum of periods by forcing the Ledoux term $B$ to 
be zero locally at the different chemical transition regions to estimate  
the impact that each single interface has on the mode-trapping properties 
\citep[see, for instance,][]{1992ApJS...81..747B,2005A&A...439L..31C}. 
This simple exercise reveals that the strong minima in $\Delta \Pi$ for periods longer 
than $\sim 900-1000$ s in our new models are due to the presence of the very sharp 
peak in $N$  near the stellar centre, which originates from the abrupt step of the 
chemical composition profile in the core region. This feature is absent in the 
R12 models. We show in Fig. \ref{x-b-n2-bvariable} the chemical profile of O 
(upper panel) at the O/C interface for a model with $M_{\star}= 0.5932 M_{\odot}$ and 
$T_{\rm eff}\sim 12\,000$ K in the case of the present model computations and in the case 
of the R12 model set. The difference in the slope of the drop in the O abundance 
for our model and that of the R12 models is very noticeable (see around $\log(1-M_r/M_{\star})\sim -0.32$). 
We also show the Ledoux term $B$ at the region of the O/C interface of our model 
(middle panel),  $B_{\rm OC}$, along with the situations in which we assume $B_{\rm OC}/2$, $B_{\rm OC}/4$, and the extreme case in which $B_{\rm OC}= 0$. 
Finally, we show the logarithm of the squared Brunt-V\"ais\"al\"a 
frequency (lower panel) for the same cases considered in the middle panel. 
In Fig. \ref{b_variable} we display the period spacing for the same models 
considered in Fig. \ref{x-b-n2-bvariable}. As we artificially reduce the magnitude 
of the Ledoux term, the contribution of the O/C interface becomes less and less 
important, until it is negligible in the most extreme case in which $B_{\rm OC} = 0$. 
In parallel, the very pronounced minima in the period spacings (left panel) 
become less pronounced, until they almost disappear when $B_{\rm OC}= 0$. 
Note that the minima of $\Delta \Pi$ correspond to local maxima in oscillation 
kinetic-energy distribution, as depicted in the right panel. This reveals 
that these modes have finite amplitudes of oscillation in central regions, where 
the density is high, and therefore they are characterized by large kinetic energies. 
These modes can be considered as modes partially "confined" to the core of the star.
We conclude that the pronounced minima in $\Delta \Pi$ for periods longer than $\sim 900-1000$ s are produced by the sharp structure of the O/C transition zone of the core. On the other hand, the mode-trapping features for periods shorter 
than $\sim 900-1000$ s are inflicted mainly by the He/H transition region and, to a less extent, by the O/C/He interface. This is valid for both sets of models. 
We conclude that the mode-trapping characteristics of our models and those of R12, in both cases with thick H envelopes, are qualitatively similar to each other for short periods, but that they differ significantly for longer periods. 

In closing this section, we quantitatively assess the differences in the individual 
periods between our models and those of R12. In Fig. \ref{deltaperiod} we show 
the difference $\delta \Pi_k= \Pi_k- \Pi_{k,\rm R12}$ versus the radial order 
($k$) for the models examined in the previous figures. Clearly, there exist very important differences (of up to $\sim 45$ s) in excess or in defect in the pulsation periods of $g$ 
modes with the same $k$, except for the case of low-order modes, for which the differences
are lower, but still important ($\sim 20$s). The differences found in the periods and 
in the period spacings considering our new set of DA WD models have origin in the various 
improvements in the treatments of different physical processes during the evolution of the progenitors (for instance, the treatment of core helium burning) and as well as 
along the evolution of the WD stage (such as incorporating Ne diffusion and Coulomb sedimentation). This is an important result that suggests that there would be substantial 
differences in the asteroseismological determinations of ZZ Ceti stars with thick H envelopes depending on whether the old R12 models or the new models presented in this work are employed. 
This important issue will be addressed in subsequent papers (C\'orsico \& Althaus 
2021, 2022, in prep.).

\subsection{DA WDs with thin H envelopes}
\label{thin}

We have repeated the analysis of the previous section, this time 
by considering template DA WD models with thin H envelopes, 
$\log(M_{\rm H}/M_{\star}) \sim -9.3$. These envelopes are
four order of magnitude thinner than those considered in the previous section. The asteroseismological analysis of R12 indicates that many ZZ Cetis could harbour H 
envelopes this thin, so, this case is of utmost relevance. 
We show in Fig. \ref{b_ledoux-thin} the Ledoux term $B$ for our 
template models and for the models of R12, and  in Fig. \ref{bvf-thin} the corresponding 
Brunt-V\"ais\"al\"a frequency profiles. Since the chemical structure of the core 
and the triple chemical transition regions do not change in comparison with the thick H envelope models discussed above, the only novel feature of these figures is the value of $B$ at the He/H transition. We note that the value of $B$ in this external chemical transition is almost identical in our models and in the R12 models, and so is the contribution to the Brunt-V\"ais\"al\"a frequency. Therefore, the differences in the periods of both sets of models will come from the differences in the chemical structure of the core and the triple transition region. We display in Fig. \ref{period_spacing-thin} the period spacing $\Delta \Pi$ for the models with thin H envelope considered in Figs. \ref{b_ledoux-thin} and \ref{bvf-thin}. As in the case of thick H envelope models, 
we note that the period spacing is qualitatively similar for both sets of models for periods shorter than $\sim 900-1000$ s, but that important differences appear for longer periods. Specifically, the R12 models show a gradual approach of $\Delta \Pi$ to the asymptotic period spacing, while the models presented in this work exhibit strong minima for increasing periods. We have verified that, as in the case of thick H envelopes (see previous section), is the presence of the very sharp 
peak in $N$  near the stellar centre of our 
models the responsible for the strong minima in $\Delta \Pi$ for long periods. 
This is at variance with the R12 models. The He/H transition, on the other hand, plays a secondary role for long periods, although it has a strong impact on $\Delta \Pi$ for short and intermediate periods. This last  property is also typical of the R12 models. To quantify the discrepancies in the individual periods between our models and those of R12, we depict in Fig. \ref{deltaperiod-thin} the differences of our periods minus those of R12 for fixed radial order. We conclude that period differences among  
$\sim 20$ s and $\sim 40$ s that we find between both sets of calculations, could affect the asteroseismologically-derived stellar parameters of ZZ Ceti stars, 
in particular for long-period pulsators, even in the case of thin H envelopes.

\subsection{DA WDs with ultra-thin H envelopes}
\label{ultra-thin}

Finally, we describe here the pulsation properties of DA WD models with ultra-thin H 
envelopes, that is, $\log(M_{\rm H}/M_{\star}) \sim -13.5$. 
Since there are no DA WD models of R12 with such thin H 
envelopes\footnote{They reach, at most, up to $\log(M_{\rm H}/M_{\star}) \sim -9.3$.}, 
we are prevented from comparing our results with their computations. 
The comprehensive survey of DA WD models of \cite{1992ApJS...81..747B} includes 
H envelopes as thin as $\log(M_{\rm H}/M_{\star}) \sim -14$. However, since those 
old models have cores made of pure C with no O, they are not directly comparable 
to ours, which harbour C/O cores. We show in Fig. \ref{b_ledoux-ultra-thin} the Ledoux 
term  $B$ and in Fig. \ref{bvf-ultra-thin} the logarithm of the squared 
Brunt-V\"ais\"al\"a frequency for template models with ultra-thin H envelopes. 
We note that for the model with $M_{\star}= 0.8373 M_{\odot}$, the 
peak corresponding to the He/H interface is very sharp and high compared to 
the models of the other masses. This is because, for such a thin H envelope and for that
stellar mass, a thin convective region appears in the He/H transition zone that mixes He and H, generating an abrupt step in the chemical profile that gives rise to that sharp peak in $B$. The resulting period-spacing diagrams are depicted in Fig. \ref{period_spacing-ultra-thin}. These diagrams look very similar to the previous ones corresponding to thicker envelopes, i.e. a pattern of minima and maxima more or less irregular, due to the impact of the He/H transition and the triple O/C/He transition, for periods shorter than $\sim 900-1000$ s, and a quite regular pattern characterized by very deep minima due mainly to the chemical transition of the CO core, for periods longer than $\sim 900-1000$ s.

The inclusion of ultra-thin H envelopes in realistic models of DA WDs opens up for the first time the possibility of exploring the internal structure of ZZ Ceti stars with such thin H envelopes with asteroseismology.

\section{Summary and conclusions}
\label{conclusions}

In this paper, we have presented a new grid of DA WD models appropriate for asteroseismology 
of ZZ Ceti stars that incorporates the advances in the last decade in the modeling and input physics of both the WD progenitors and WD stars.  This new grid of models constitutes a major improvement over the ZZ Ceti models developed by \cite{2010ApJ...717..897A} and \cite{2010ApJ...717..183R}, which we have  amply used in numerous asteroseismological analyses like the study of R12 and subsequent works. Specifically, we  generated DA WD stellar models with masses from $\sim 0.52$ to  $\sim 0.83 M_{\sun}$, resulting from the whole evolution of initially 1.5 to  4.0 $M_{\sun}$ mass-star models, which embraces the range of stellar masses expected for most of the observed ZZ Ceti stars. Our models were derived in a self-consistent way with the changes in the internal chemical composition due to the mixing of core constituents induced by mean molecular weight inversions,  $^{22}$Ne diffusion,  Coulomb sedimentation, and  residual nuclear burning. Our new models also accounts for nuclear burning history and mixing events along the progenitor evolution, in particular the occurrence of third dredge-up, which determine the properties of the core and envelope of post-AGB and WD stars, as well as the WD initial-final mass relation. The range of H envelope thickness of our models extends from the maximun residual H content predicted by progenitor history, $\log(M_{\rm H}/M_{\sun})\sim -4$ to $-5$, to  $\log(M_{\rm H}/M_{\odot})= -13.5$, thus allowing for 
the first time to make available models that will allow in the short term 
to find seismological solutions for ZZ Ceti stars with extremely thin H envelopes, if they exist in Nature. Calculations have been performed with the {\tt LPCODE} stellar 
evolution code.

Our  new H-burning post-AGB models  predict chemical structures for the ZZ Ceti stars substantially different from those used in seismological studies by R12. We discussed the implication of these new models for  the pulsational spectrum of ZZ Ceti stars. 
We find that the pulsation periods of  $g$ modes and mode trapping properties  of 
our new models differ significantly from those characterizing the R12's ZZ Ceti models. Specifically, our new models are characterized by a very abrupt step in the C and O 
abundances at the  stellar core, which have a dramatic impact on the pulsation periods themselves and also on the mode-trapping properties and consequently the period spacing. 
In the case of the period spacing, our models and those of R12 agree quite well for periods shorter than $\sim 900-1000$ s, but strongly differ for longer periods. Specifically, for high radial-order modes ($\Pi \gtrsim 900-1000$ s), our models show a distribution of period spacings characterized by very pronounced minima, related to high oscillation-energy modes that have finite amplitudes in the CO-core regions. Instead, the period spacing of the R12 models gradually approaches the asymptotic period spacing for long periods. 
On the other hand, the comparison between the periods of our models and the ones of the R12 models indicates differences from $\sim 20$ s to $\sim 45$ s for modes of equal radial order, calculated on models of WDs with equal mass, effective temperature and H-layer thickness. The differences in the periods and period spacings are due to several improvements we have made in 
modeling the evolution of WD progenitors (for instance, the treatment of 
core He burning), as well as in the evolution during the WD stage, such as the Ne diffusion and the Coulomb sedimentation. These differences in the periods and period spacings likely have the important consequence that the parameters of ZZ Ceti stars derived through asteroseismological period-to-period fits up to date could substantially vary if 
our new set of models were used. Furthermore, our models include extremely thin 
H envelopes, $\log(M_{\rm H}/M_{\odot})= -13.5$, ten thousand times thinner than the  thinnest H envelopes considered in R12, $\log(M_{\rm H}/M_{\odot})\sim -9.3$. This 
improvement paves the way for revisiting the asteroseismological determinations made so far 
considering the possibility that some ZZ Cetis can harbour ultra-thin H envelopes.

We close the paper by emphasizing that the new high-quality photometric observations 
provided by already completed space missions like {\sl Kepler/K2} \citep[][]{2010Sci...327..977B, 2014PASP..126..398H}, 
ongoing space programs such as {\sl TESS} \citep[][]{2015JATIS...1a4003R} and {\sl Cheops} \citep[][]{2018A&A...620A.203M}, 
and future space telescopes like {\sl Plato} \citep[][]{2018EPSC...12..969P}, pose 
a challenge to WD modellers, who must substantially improve stellar models 
to be used in high-precision WD asteroseismology. The development of a new generation 
of detailed evolutionary models of DA WDs to interpret the new observations constitutes the first step in that direction. This has been precisely the objective of this paper. 
Our future works will be focused on the calculation of a huge grid of stellar models and their adiabatic $g$-mode periods, and their application in asteroseismological analyses of the ZZ Cetis already known, and also of the new ones that will be discovered in the near future.

\begin{acknowledgements}
 
We  acknowledge  the  suggestions  and comments of an 
anonymous referee that improved the original version of this paper.  
Part of  this work was  supported by PICT-2017-0884 from ANPCyT, PIP
112-200801-00940 grant from CONICET, grant G149 from University of La Plata.
This  research has  made use of  NASA Astrophysics Data System.
\end{acknowledgements}

\bibliographystyle{aa}
\bibliography{bibliografia}



\end{document}